\documentclass[twocolumn,floats,floatfix,aps,pra]{revtex4-2}
\usepackage{amsfonts,amssymb,amsmath}
\usepackage{color,calc}
\usepackage[dvips]{graphicx}
\usepackage{bm}
\usepackage{multirow}

\def\be{ \begin{equation} }
\def\ee{ \end{equation} }
\def\bea{ \begin{eqnarray} }
\def\eea{ \end{eqnarray} }
\def\bse{ \begin{subequations} }
\def\ese{ \end{subequations} }
\def\ba{ \begin{array} }
\def\ea{ \end{array} }

\def\i{\,\text{i}}

\def\i{i}

\def\to{\rightarrow}

\def\T12{T_{\sfrac{1}{2}}}

\def\U{\mathbf{U}}

\def\R{\mathbf{R}}
\def\G{\mathbf{G}}

\newcommand{\ket}[1]{\vert #1\rangle}
\def\i{{\rm{i}}}
\def\f{{\rm{f}}}

\def\i{\textrm{i}}
\def\f{\textrm{f}}

\def\i{\rm{i}}
\def\f{\rm{f}}
\def\r{\rm{r}}
\def\rottheta{\vartheta}
\def\ai{\rm ai}
\def\width{\tau}
\def\ts{\lambda}
\def\etaLMSZ{\eta'}

\begin{document}

%%%%%%%%%%%%%%%%%%%%%%%%%%%%%%%%%%%%%%%%%%%%%%%%%%%%%%%%%%%%%%%%%%%%%%%%%%%%%%%%%%%%%%%%%%%%%%%%%%%%%%%%%%%%%%%%%%%%%%%%%%%%%%%%%%%%%%%%%
%%% TITLE PAGE %%%

\title{Qubit dynamics driven by smooth pulses of finite duration}

\author{Ivo S. Mihov and Nikolay V. Vitanov}

\affiliation{Center for Quantum Technologies,  Department of Physics, St Kliment Ohridski University of Sofia, 5 James Bourchier blvd, 1164 Sofia, Bulgaria}

\date{\today }

\begin{abstract}
We present a study of the dynamics of a qubit driven by a pulsed field of finite duration. 
The pulse shape starts and ends linearly in time.
The most typical example of such a shape is the sine function between two of its nodes, but several other pulse shapes are also studied.
All of them present smooth alternatives to the commonly used rectangular pulse shape, resulting in much weaker power broadening, much faster vanishing wings in the excitation line profile and hence much reduced sidebands.
In the same time, such shapes with a well-defined finite duration do not suffer from the spurious effects arising when truncating a pulse of infinite duration, e.g. Gaussian.
We derive two approximate analytic solutions which describe the ensuing quantum dynamics.
Both approximations assume that the field changes linearly at the beginning and the end of the driving pulse, and adiabatically in between. 
The first approximation matches the linear and adiabatic parts at an appropriate instant of time and is expressed in terms of Weber's parabolic cylinder functions.
The second, much simpler, approximation uses the asymptotics of the Weber function in order to replace it by simpler functions, and some additional transformations.
Both approximations prove highly accurate when compared to %numerical solutions to the Schr\"odinger equation. 
experimental data obtained with two of the IBM Quantum processors.
Both the greatly reduced power broadening and the greatly suppressed sidebands are observed for all pulse shapes, in a nearly complete agreement between theory and experiment. 
\end{abstract}

\maketitle

%%%%%%%%%%%%%%%%%%%%%%%%%%%%%%%%%%%%%%%%%%%%%%%%%%%%%%%%%%%%%%%%%%%%%%%%%%%%%%%%%%%%%%%%%%%%%%%%%%%%%%%%%%%%%%%%%%%%%%%%%%%%%%%%%%%%%%%%%
\section{Introduction\label{Sec:derivation1}}
%%%%%%%%%%%%%%%%%%%%%%%%%%%%%%%%%%%%%%%%%%%%%%%%%%%%%%%%%%%%%%%%%%%%%%%%%%%%%%%%%%%%%%%%%%%%%%%%%%%%%%%%%%%%%%%%%%%%%%%%%%%%%%%%%%%%%%%%%

The profound understanding and control of qubit dynamics is increasingly important in the quest for a scalable fault-tolerant quantum computer, where one of the key steps is achieving high-fidelity gate operations.
Quantum gates are implemented by electromagnetic pulses, the shape of which has a significant impact on the gate properties.
The pulse shape has been recognized as an important control parameter in numerous applications of quantum control, including
STIRAP \cite{Vasilev2009, Du2016, Stefanatos2021, Zheng2022}, 
quantum memories \cite{Heinze2013, Wenner2014},
quantum optics \cite{Vasilev2010,Ansari2018,Dais2019}, 
spectroscopy and sensing \cite{Zopes2017, Keefer2021},
ultrafast effects \cite{Bayer2009},
optimization of quantum gates \cite{Zhu2006, Campbell2010, Hayes2012, Daems2013, Choi2014, Schafer2018, Zarantonello2019, Rimbach-Russ2023, Xie2023},
including suppression of leakage \cite{Motzoi2009, Chen2016, Werninghaus2021}, to name just a few.

The most popular pulse shapes used for qubit control are the rectangular and Gaussian pulses and in superconducting qubits, the DRAG pulse~\cite{Motzoi2009}. 
The rectangular pulse packs most pulse area in the shortest time duration but it suffers from excitation sidebands and power broadening which may lead to detrimental cross talk to other states. 
The Gaussian pulse assumes an infinite duration and has to be truncated in practical use; the truncation can cause some artefacts in its excitation profile and in order to avoid these one has to move the truncation times further away from its center thereby making it too long and the gate too slow.
Similarly, truncation is needed in the case of the commonly used DRAG pulses~\cite{Motzoi2009, Gambetta2011}.
They are the state-of-the-art when working with superconducting qubits, where the anharmonicity (the detuning from the transition to the third energy level) is only 5-10\% of the pulse's Rabi frequency.

To this end, the analytic description of qubit dynamics is an invaluable tool because, contrary to numerical optimizations, it provides qualitative understanding of the role and significance of various experimental parameters. 
In this paper, we consider a very important class of pulse shapes, which involves \textit{smooth} pulses of well-defined \textit{finite} duration, which start and end at zero with a linear slope.
A typical representative of this class is the sine pulse. 
Other examples include a variety of bell-shaped smooth pulses which are truncated and their ensuing background is subtracted: Gaussian, hyperbolic-secant and its powers, Lorentzian and its powers, etc.

This class of pulses have some clear advantages. 
Compared to rectangular pulses, they are smooth, with no sudden amplitude jumps in the beginning and the end, meaning no large sidebands in the excitation profile and less power broadening.
Compared to bell-shaped pulses (e.g. Gaussian, hyperbolic-secant, Lorentzian, etc.), they are finite and there is no need to truncate their wings (and compensate for the ensuing adverse effects).
Compared to these latter pulses, the present type of pulses offers significant ``filling ratio'', i.e. large pulse area confined in a short duration. 
For instance, for a given pulse area, the sine pulse requires a pulse duration which is a factor of only $\pi/2$ longer than that for a rectangular pulse, while mitigating the drawbacks of the latter pulse.
Therefore these smooth pulses of finite duration can be considered as candidates for the ``sweet spot'' of qubit control.

In order to describe and understand the qubit dynamics when driven by such shapes, we need some analytic description.
Unfortunately, none of the existing exact or approximate methods are applicable to such pulse shapes.
There are several exactly soluble analytic models for the coherent interaction of a two-state quantum system with an external field, 
e.g. the models of Rabi \cite{Rabi1937}, Rosen-Zener~\cite{Rosen1932}, Allen-Eberly-Hioe \cite{Allen1975,Hioe1981}, Demkov-Kunike \cite{Demkov1969}, Landau-Majorana-St\"uckelberg-Zener (LMSZ)~\cite{Landau1932,Zener1932,Stueckelberg1932,Majorana1932}, Demkov~\cite{Demkov1963}, Carroll-Hioe \cite{Hioe1985}, etc.
All these models, except for the ones with a rectangular pulse shape (the Rabi model and the finite LMSZ model), assume an infinite pulse duration, e.g. hyperbolic secant or exponential.
In addition, there are several approximate methods, such as the perturbation theory, the Magnus approximation \cite{Blanes2009}, the Davis-Pechukas approach \cite{Davis1976}, the Rosen-Zener conjecture \cite{Rosen1932,Conover2011,Mihov2023}, etc., which assume different limitations and deliver different accuracy. 
The first two of these work in the weak-field regime and are not suitable for qubit operations.
The beautiful and powerful Davis-Pechukas approach demands analytic pulse shapes (implying infinite pulse duration).
The Rosen-Zener conjecture is an interesting but rarely used proposition, which has never been rigourously proved.

Pulse shaping techniques have lately been found to provide new approaches to old issues in quantum control. 
Lorentzian pulses have been demonstrated as a treatment for power broadening, and other pulse shapes have also been found to provide reduced power broadening, in comparison with the rectangular pulse~\cite{Mihov2024, Vitanov2001}.
Trying to balance between these extreme cases, the sinusoidal pulse shape provides a valuable option for uses where the duration of the pulse is important.
They have been the object of many scientific works in the recent years~\cite{Patel2020, Zopes2017, Datta2022, Xu2023, Boradjiev2013}.
Unfortunately, these applications could not make use of an analytic solution for the sinusoidal pulse shape because the sine model is not exactly soluble.

To this end, in order to describe the qubit dynamics we derive here two highly accurate analytic approximations which are especially designed (and applicable) for such pulse shapes.
They are based on the assumption that the qubit dynamics in the beginning and the end of the pulse is described by the rotated half-crossing LMSZ model, which is applicable for linearly changing pulse shapes.
In the vicinity of the pulse maximum the evolution is assumed to be adiabatic.
We derive two approximate formulae, one using the exact solution for the rotated LMSZ model, which involves Weber's parabolic cylinder function, and another using approximations to it in terms of simpler functions.
The latter approximation turns out to be very simple and, to our surprise, very accurate.
We test these two approximations on a variety of pulse shapes by comparing them to data acquired on one of the quantum processors of IBM Quantum.

We note that some features of excitation produced by the sine pulse have been considered in Ref.~\cite{Boradjiev2013}, namely the line width $\Delta_{\frac12}$ of the excitation profile. 
The approach to estimate the line width used the nonanalyticities at the beginning and the end of the pulse.
It predicted power broadening with the square root of the peak Rabi frequency, $\Delta_{\frac12} \propto \Omega_0^{\frac12}$, and an excitation profile vanishing as $\Delta^{-4}$ for large detuning $\Delta$. 
The approach is only valid for large detunings and large Rabi frequencies and fails to correctly predict the entire line profile, e.g. at resonance and small detunings. 
Here we provide a complete description of the excitation by such pulses, which is valid for any values of the experimental parameters, as long as the pulse shape obeys some restrictions.

The paper is organized as follows.
Section \ref{Sec:general} introduces the problem and the main assumptions made, and
Secs. \ref{Sec:derivation1} and \ref{Sec:derivation2} present the two approaches for derivation of the analytic approximations. 
Section \ref{Sec:demo} shows experimental data and comparisons to the analytic approximations, 
Sec. \ref{Sec:applicability} discusses the conditions of validity of the approximations, and Sec.~\ref{Sec:conclusion} summarizes the conclusions.

%%%%%%%%%%%%%%%%%%%%%%%%%%%%%%%%%%%%%%%%%%%%%%%%%%%%%%%%%%%%%%%%%%%%%%%%%%%%%%%%%%%%%%%%%%%%%%%%%%%%%%%%%%%%%%%%%%%%%%%%%%%%%%%%%%%%%%%%%
\section{General\label{Sec:general}}
%%%%%%%%%%%%%%%%%%%%%%%%%%%%%%%%%%%%%%%%%%%%%%%%%%%%%%%%%%%%%%%%%%%%%%%%%%%%%%%%%%%%%%%%%%%%%%%%%%%%%%%%%%%%%%%%%%%%%%%%%%%%%%%%%%%%%%%%%

In this paper, we consider a class of driving pulses with constant detuning and smooth temporal shapes, which (i) have a finite duration $T$, (ii) are symmetric in time, (iii) start from zero and end to zero with a nonzero derivative,
\bse
\begin{align}
\Omega(t) &= \left\{
\begin{array}{cc}
\Omega_0 f(t), & 0\leqq t \leqq T, \\
0, & \text{otherwise},
\end{array}
 \right.
 \\
\Delta(t) &= \text{const},
\end{align}
\ese
where
\bse
\begin{align}
f(T-t) &=f(t),
 \\
 f(0) &= f(T) = 0,
 \\
 f'(0) &= -f'(T) > 0.
\end{align}
\ese
and (iv) their initial and final segments are approximately linear in time, i.e.

\be
    f'(t) \sim \text{const for } t \ll T/2 \text{ and } t-T \ll  T/2.
\ee
%***************************************************************
\begin{figure}[t]
\includegraphics[width=0.95\columnwidth]{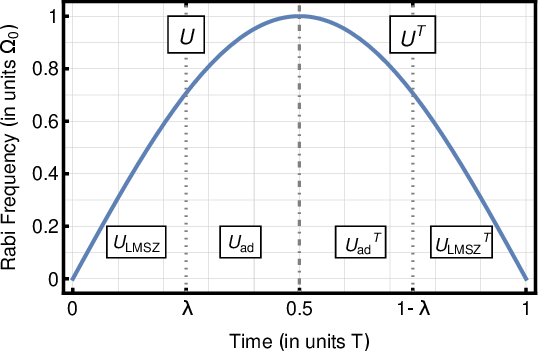}
\caption{
Illustration of the assumptions used in the derivation of the split LMSZ-adiabatic approximation which assumes rotated-LMSZ evolution at $t \in [0, \ts]\cup [1-\ts,1]$ of the driving pulse and adiabatic evolution for $t \in [\ts,1-\ts]$.
}
\label{fig-sine}
\end{figure}
%***************************************************************

We shall derive two analytic approximations to the solution of the Schr\"odinger equation for such pulse shapes in the most general case when only the above properties are imposed and no other restriction on the specific shape are assumed.
In both approximations, we first separate the evolution into two intervals: $[0,\frac12 T]$ and $[\frac12 T,T]$.
Because the pulse shape is assumed to be symmetric in time, it can be shown that if the propagator for the first half-interval is $\U$, then the propagator for the second half-interval is $\U^T$.
Therefore, the overall propagator in the entire interval is $\U^T \U$ and it suffices to consider the evolution in the first half interval $[0,\frac12 T]$ only.

In the second step of both methods, we first rotate the basis at an angle $\pi/4$, described by the matrix (which amounts to the Hadamard gate)
\be
\G = \tfrac{1}{\sqrt{2}}\left[ \begin{array}{cc} -1 & 1 \\ 1 & 1 \end{array} \right] ,
\ee
so that the Rabi frequency and the detuning in the initial Hamiltonian,
\be\label{Horig}
\mathbf{H}(t) = \tfrac12 \left[ \begin{array}{cc} -\Delta(t) & \Omega(t) \\ \Omega(t) & \Delta(t) \end{array} \right],
\ee
exchange their places in the rotated Hamiltonian,
\be\label{Hrot}
\widetilde{\mathbf{H}}(t) = 
\G \mathbf{H}(t) \G
= \tfrac12 \left[ \begin{array}{cc} -\Omega(t) & \Delta(t) \\ \Delta(t) & \Omega(t) \end{array} \right] .
\ee
Note that $\G^{-1}=\G$ and hence $\G^2 = \mathbf{1} $.
If the propagator for the original Hamiltonian is $\U(t_{\f},t_{\i})$ then the propagator for the rotated Hamiltonian reads
\be
\widetilde{\U}(t_{\f},t_{\i}) = 
\G \U(t_{\f},t_{\i}) \G .
\ee

In this rotated picture, the problem becomes one with a constant coupling $\Delta$ and a pulse-shaped detuning $\Omega(t)$. 
This allows us to model the evolution as two LMSZ half-crossings at the beginning and at the end of the pulse, and adiabatic evolution in between.
Both assumptions --- the linear behavior in the two ends and the adiabatic evolution in between --- impose some restrictions on the experimental parameters which will be discussed in due course.

In the first approximation, it is assumed that the rot-LMSZ part lasts in the interval $[0,\ts]$, and the adiabatic part is in the interval $[\ts,\frac12 T]$ (see Fig.~\ref{fig-sine}), where $\ts$ is a free fitting parameter, which depends on the pulse shape.
The rot-LMSZ propagator is expressed in terms of Weber's parabolic cylinder function, whereas the adiabatic solution is given in terms of trigonometric functions and an adiabatic phase.

The second approximation is more sophisticated as it intermixes the rot-LMSZ and adiabatic solutions in the adiabatic interaction representation, uses the so-called strong-coupling asymptotics of the Weber function and replaces the rot-LMSZ dynamic phase with the adiabatic phase.
The result is a much simpler analytic formula, which is expressed in terms of elementary functions, the argument of the Euler's Gamma function, and the adiabatic phase.
Moreover, this much simpler analytic formula, which does not include any fitting parameters, fits the exact solution equally well as the first approximation.

%%%%%%%%%%%%%%%%%%%%%%%%%%%%%%%%%%%%%%%%%%%%%%%%%%%%%%%%%%%%%%%%%%%%%%%%%%%%%%%%%%%%%%%%%%%%%%%%%%%%%%%%%%%%%%%%%%%%%%%%%%%%%%%%%%%%%%%%%
\section{First method: Split LMSZ-adiabatic approximation\label{Sec:derivation1}}
%%%%%%%%%%%%%%%%%%%%%%%%%%%%%%%%%%%%%%%%%%%%%%%%%%%%%%%%%%%%%%%%%%%%%%%%%%%%%%%%%%%%%%%%%%%%%%%%%%%%%%%%%%%%%%%%%%%%%%%%%%%%%%%%%%%%%%%%%

In the first method, we separate the first half interval $[0,\frac12 T]$ into two further sub-intervals: $[0,\ts ]$ and $[\ts,\frac12 T]$, see Fig.~\ref{fig-sine}, where $\ts$ is a free fitting parameter.
In the first sub-interval $[0,\ts ]$, we approximate the solution by the rot-LMSZ propagator $U_{\text{LMSZ}}$ due to the near linearity of the pulse shape near $t=0$.
Obviously, the smaller the parameter $\ts$, the better the linear assumption and the better the rot-LMSZ approximation.
In the second sub-interval $[\ts,\frac12 T]$, we approximate the solution by the adiabatic solution.
This choice is justified if in this interval $\Omega(t)$ is sufficiently large (meaning that its maximum $\Omega_0$ is large enough).

The rot-LMSZ propagator in the first sub-interval $[0,\ts ]$ is already available in the literature \cite{Torosov2008}; it reads
\be
\U_{\text{LMSZ}} = \left[ \begin{array}{cc} \Re a-i\Im b & \Re b + i\Im a \\ -\Re b + i\Im a & \Re a +i\Im b \end{array} \right],
\ee
with
\bse\label{ab-LMSZ}
\begin{align}
a &=
\frac{2^{\frac12 i \delta ^2} (1+e^{-\pi  \delta ^2}) \Gamma \left(\frac12+\frac12{i \delta^2}\right)}{2\sqrt{\pi }} D_{-i \delta ^2}(\alpha  e^{\frac{i \pi }{4}})
\notag\\
 &+ \frac{ (1-e^{-\pi  \delta ^2}) \Gamma \left(1-\frac12{i \delta^2}\right)}{2^{\frac12i \delta^2} \sqrt{2\pi }} D_{i \delta ^2-1}(\alpha  e^{-\frac{i \pi }{4}}) ,
\\
b &=
\frac{e^{-i\pi/4} 2^{\frac12i \delta^2} (1-e^{-\pi  \delta ^2}) \Gamma (1+\frac12{i \delta ^2})}{\delta \sqrt{2 \pi } }
 D_{-i \delta ^2} (\alpha  e^{\frac{i \pi}{4}})
  \notag \\
&- \frac{e^{-i\pi/4} (1+e^{-\pi  \delta ^2}) \delta \Gamma (\frac12-\frac12{i \delta ^2})}{2^{1+\frac12i \delta^2} \sqrt{ \pi } }
 D_{i \delta ^2-1} (\alpha  e^{-\frac{i \pi}{4}}),
\end{align}
\ese
where $D_\nu(z)$ is Weber's parabolic cylinder function \cite{Wolfram-Weber}, $\Gamma(z)$ is Euler's gamma function \cite{Wolfram-gamma}, and
\be\label{LMSZparams}
\alpha = \beta \ts, \quad 
\delta = \frac{\Delta}{2\beta}, \quad 
\beta = \sqrt{\Omega_0 f(\ts)}.
\ee

The adiabatic solution for the second sub-interval $[\ts,\frac12 T]$ reads
\bse\label{Uad}
\be
\U_{\text{adb}} = \left[ \begin{array}{cc} c & d \\ -d^* & c^* \end{array} \right],
\ee
where
\begin{align}
c &= e^{-\frac{1}{2} i \eta_{\ts}} \sin \theta _{\i} \sin \theta _{\f} + e^{\frac{1}{2} i \eta_{\ts}}  \cos \theta _{\i} \cos \theta _{\f} ,
\\
d &= e^{-\frac{1}{2} i \eta_{\ts}}  \cos \theta _{\i} \sin \theta _{\f} - e^{\frac{1}{2} i \eta_{\ts}} \sin \theta _{\i} \cos \theta _{\f}.
\end{align}
\ese
Here the mixing angle $\theta(t)$ is defined by 
\be\label{theta}
\cos2\theta(t) = \frac{\Delta }{\Lambda(t)},
\quad
\sin2\theta(t) =  \frac{\Omega(t)} %0 \sin (\pi  t/T )}
{\Lambda(t)},
\ee
and $\theta_{\i}=\theta(t_{\i})=\theta(\ts)$ and $\theta_{\f}=\theta(t_{\f})=\theta(\frac12T)$.
The eigenfrequency splitting $\Lambda(t)$ and the adiabatic phase $\eta_{\ts}$ read
\bse
\begin{align}
\label{Lambda}
\Lambda(t) &= \sqrt{\Delta ^2+ \Omega(t)^2}, 
\\
\label{eta}
\eta_{\ts} &= \int_{\ts}^{\frac12T} \Lambda(t) \, dt .
\end{align}
\ese
Most quantities above --- the LMSZ parameters $\alpha$, $\beta$, $\delta$, the mixing angle $\theta_{\i}$, and the adiabatic phase $\eta$ --- depend on the value of the point $\ts$, which separates the LMSZ and adiabatic intervals.
It looks plausible to choose $\ts=\frac14T$, but the accuracy of the formula can be improved if it is left as a free fitting parameter, which has a different value for each pulse shape.

The propagator for the half interval $[0,\frac12 T]$ is $\U_{\text{adb}} \U_{\text{LMSZ}} $, and the propagator for the full interval $[0,T]$ is
\be
\label{UrLMSZ}
\U =
\U_{\text{LMSZ}}^T \U_{\text{adb}}^T \U_{\text{adb}} \U_{\text{LMSZ}}.
\ee
The two adjacent adiabatic propagators can be combined into a single adiabatic propagator, which describes the evolution in the time interval $[\ts, T-\ts]$,
which can be obtained from Eq.~\eqref{Uad} by setting $\theta _{\f}=\theta _{\i}$ and $\eta_{\ts}\to 2\eta_{\ts}$.
The transition probability for the full interval $[0,T]$ can be obtained from the overall propagator \eqref{UrLMSZ} as 
$P=|U_{12}|^2$; explicitly,
\begin{align}
P =& \left|
\Re [ (a^2-b^2) \sin 2 \theta_{\i}  
+ 2 a b^* \cos 2\theta _{\i} ] \sin \eta_{\ts}
\right. \notag \\ 
&+ \left. 
 \Im(a^2-b^2) \cos \eta_{\ts}
\right|^2 .
\label{eq-Psplit}
\end{align}
A much simpler expression for the transition probability is obtained by the second method in the next section.

%%%%%%%%%%%%%%%%%%%%%%%%%%%%%%%%%%%%%%%%%%%%%%%%%%%%%%%%%%%%%%%%%%%%%%%%%%%%%%%%%%%%%%%%%%%%%%%%%%%%%%%%%%%%%%%%%%%%%%%%%%%%%%%%%%%%%%%%%
\section{Second method: Integrated LMSZ-adiabatic approximation\label{Sec:derivation2}}
%%%%%%%%%%%%%%%%%%%%%%%%%%%%%%%%%%%%%%%%%%%%%%%%%%%%%%%%%%%%%%%%%%%%%%%%%%%%%%%%%%%%%%%%%%%%%%%%%%%%%%%%%%%%%%%%%%%%%%%%%%%%%%%%%%%%%%%%%

For the derivation of the second analytic formula we follow an approach similar to the one in Appendix D of Ref.~\cite{Garraway1997}.
As noted above, it suffices to consider the evolution in the first half interval $[0,\frac12 T]$.
Now we do not split this half interval further, as in the first approach, but treat it as a whole. 
We parameterize the propagator for the rotated Hamiltonian \eqref{Hrot} in this interval as 
\be\label{Urot}
\widetilde{\U}(\tfrac12 T,0) = \left[ \begin{array}{cc} u & v \\ -v^* & u^* \end{array} \right].
\ee
Then the overall propagator in the original (non-rotated) basis in the \textit{entire} interval $[0,T]$ is expressed as
\be
\U(T,0) = 
\G \widetilde{\U}(\tfrac12 T,0)^T \widetilde{\U}(\tfrac12 T,0) \G 
= \left[ \begin{array}{cc} x & y \\ -y^* & x^* \end{array} \right],
\ee
 with
\bse
\begin{align}
x &= \frac{(u^*+v^*)^2+(u-v)^2}{2}  
 = (u_{\r} - i v_{\i})^2 + (v_{\r} - i u_{\i})^2, \\
y &= \frac{u^{*2} - u^2 + v^2 - v^{*2}}{2}  
= -2 i (u_{\r} u_{\i} - v_{\r} v_{\i}) .
\end{align}
\ese
Here, in order to avoid confusion, we explicitly keep the arguments of the propagator.

Next we transform the rotated basis to the adiabatic interaction picture. 
The reason to use this picture is that, contrary to other bases, the propagator elements (and especially their phases) in the original LMSZ model (i.e. with infinite interaction duration) are convergent.
We do this in two steps: first, we transform the rotated basis to the adiabatic rotated basis, and second, we transform the latter to the adiabatic interaction representation. 

We begin with the rotated Hamiltonian \eqref{Hrot} 
and transform it to the adiabatic representation --- the basis of its eigenvectors,
\be\label{EVec}
\ket{-} = \left[ \begin{array}{c} \cos\rottheta(t) \\ -\sin\rottheta(t) \end{array} \right], \quad
\ket{+} = \left[ \begin{array}{c} \sin\rottheta(t) \\ \cos\rottheta(t) \end{array} \right],
\ee
where 
\be\label{rottheta}
\sin 2\rottheta(t) = \frac{\Delta} {\Lambda(t)},\quad \cos 2\rottheta(t) = \frac{\Omega(t)} {\Lambda(t)},
\ee
and the eigenvalue splitting $\Lambda(t)$ is defined in Eq.~\eqref{Lambda}.
Note that the mixing angles in the original basis $\theta$ and in the rotated basis $\rottheta$ are related as $2\rottheta = \frac12\pi - 2\theta$, see Eqs.~\eqref{theta} and \eqref{rottheta}.
We form the transformation matrix composed of the eigenvectors \eqref{EVec},
\be
\widetilde{\R}(\rottheta(t)) = \left[ \begin{array}{cc}
\ket{-} , & \ket{+} \end{array} \right].
\ee
Note that $\widetilde{\R}(\rottheta(t))^{-1} = \widetilde{\R}(\rottheta(t))^T = \widetilde{\R}(-\rottheta(t))$.
In the adiabatic rotated representation, the Hamiltonian reads $\widetilde{\mathbf{H}}_a(t) = \widetilde{\R}(-\rottheta(t)) \widetilde{\mathbf{H}}(t) \widetilde{\R}(\rottheta(t))$; explicitly,
\be
\widetilde{\mathbf{H}}_a(t) = \tfrac12 \left[ \begin{array}{cc} -\Lambda(t) & -2i \dot\rottheta(t) \\ 2i \dot\rottheta(t) & \Lambda(t) \end{array} \right],
\ee
where the overdot means $d/dt$.

Next, we perform the transformation to the adiabatic interaction representation via the transformation matrix
\be
\mathbf{F}(\eta(t)) = \left[ \begin{array}{cc}
e^{i\eta(t)/2} & 0 \\
 0 & e^{-i\eta(t)/2}
 \end{array} \right],
\ee
with the adiabatic phase
\be
\eta(t) = \int_{t_{\i}}^{t} \Lambda(t') dt',
\label{eq-IntAdPhase}
\ee
which casts the Hamiltonian into the form $\widetilde{\mathbf{H}}_{\ai}(t) = \mathbf{F}(-\eta(t)) \widetilde{\mathbf{H}}_a(t) \mathbf{F}(\eta(t))$; explicitly,
\be
\widetilde{\mathbf{H}}_{\ai}(t) = \left[ \begin{array}{cc} 0 & -i \dot\rottheta(t) e^{-i\eta(t)} \\ i \dot\rottheta(t) e^{i\eta(t)} & 0 \end{array} \right].
\ee

The propagator \eqref{Urot}, corresponding to the Hamiltonian \eqref{Hrot},
is transformed in a similar fashion and in the adiabatic interaction representation it reads
\be
\widetilde{\mathbf{U}}_{\ai}(\tfrac12T,0) = \mathbf{F}(-\eta_{\f)}) \widetilde{\R}(-\rottheta_{\f}) \widetilde{\mathbf{U}}(\tfrac12T,0) \widetilde{\R}(\rottheta_{\i}) \mathbf{F}(\eta_{\i}),
\ee
where 
 $\eta_{\i}=\eta(0)=0$, $\eta_{\f}=\eta(\tfrac12T) \equiv \eta$,
 $\rottheta_{\i}=\rottheta(0)=\frac14\pi$, $\rottheta_{\f}=\rottheta(\tfrac12T)=\arctan (\Delta/\Omega(\frac12T))/2$.
From here, we have
\be\label{Uorig}
\widetilde{\mathbf{U}}(\tfrac12T,0)
 = \widetilde{\R}(\rottheta)\mathbf{F}(\eta) \widetilde{\mathbf{U}}_{\ai}(\tfrac12T,0) 
 %\mathbf{F}(-\eta_{\i}) 
 \widetilde{\R}(-\pi/4).
\ee

Next, we make the central assumption of this method: the evolution in the vicinity of the initial time $t=0$ follows the half-crossing LMSZ model. 
The rationale behind this assumption is that, due to the nonzero derivative of the pulse at the initial (and final) time, its departure from (and arrival to) zero can be considered linear in time. 
Moreover, we assume that the essential part of the dynamics occurs in the vicinity of the half-crossing (as $t\to 0$) and very little changes away from it (as $t\to \frac12 T$).
The rationale behind this assumption is that the nonadiabatic coupling $\dot\rottheta(t)$ is maximal as $t\to 0$, because $\dot\Omega(t)$ is maximal there, whereas in the other end of the half-interval ($t\to \frac12T$) we have $\dot\Omega(t) \to 0$ and hence $\dot\rottheta(t) \to 0$.
Moreover, it is well known that the transition time in the adiabatic basis is much shorter than the transition time in the original basis \cite{Vitanov1999}, which means that the LMSZ transition can be assumed to be well localized in the vicinity of the point $t=0$.

The half-LMSZ evolution is described by the propagator (Appendix \ref{Sec:AdbInt-LMSZ})
\be\label{ULMSZ}
\widetilde{\U}_{\text{LMSZ}}= \tfrac{1}{\sqrt{2}}
\left[
\begin{array}{cc}
 \sqrt{1+e^{-\pi\delta^2}} e^{i \chi _1} & \sqrt{1-e^{-\pi\delta^2}} e^{i \chi _2} \\
 -\sqrt{1-e^{-\pi\delta^2}} e^{-i \chi _2} & \sqrt{1+e^{-\pi\delta^2}} e^{-i \chi _1}
\end{array}
\right] ,
\ee
with
%%%%%%%  NE TRII TOVA!!!!! %%%%%%%%%%%%%%%%
%\be
%\xi _1 = \chi_1 -\frac{\etaLMSZ}{2} ,\quad
%\xi _2 = \chi_2 - \frac{\etaLMSZ}{2} ,
%\ee
%where $\etaLMSZ$ is the LMSZ adiabatic phase \eqref{etaLMSZ} and
%%%%%%%  NE TRII TOVA!!!!! %%%%%%%%%%%%%%%%
\bse\label{chi1-chi2}
\begin{align}
\chi _1 &= \frac{\delta ^2}{2} - \frac{\delta ^2}{2}  \ln \left(\frac{\delta ^2}{2}\right) + \arg \Gamma \left(\frac{1+i \delta ^2}{2}\right) ,\\
\chi _2 &= \frac{\delta ^2}{2}-\frac{\delta ^2}{2}  \ln \left(\frac{\delta ^2}{2}\right) +  \arg \Gamma \left(\frac{i \delta ^2}{2}\right) + \frac{\pi }{4},
\end{align}
\ese
and
\be\label{delta}
\delta = \frac{\Delta}{2\sqrt{\pi\,\Omega\left(\frac14 T\right)}}.
\ee
The approximation consists of replacing the actual propagator $\widetilde{\mathbf{U}}_{\ai}(\frac12 T,0)$ in Eq.~\eqref{Uorig} by $\widetilde{\U}_{\text{LMSZ}}$ of Eq.~\eqref{ULMSZ}, while keeping the other matrices in Eq.~\eqref{Uorig} intact, as they are related to the actual (nonlinearized) pulse shape.
Thereby we find for the full interval $[0,T]$
\begin{align} \label{Uappr}
\mathbf{U}(T,0)
 &= 
 \G\widetilde\R(\rottheta_{\i}) %\mathbf{F}(0)^* 
 \widetilde\U_{\text{LMSZ}}^T \mathbf{F}(\eta) \widetilde\R(-\rottheta_{\f}) \notag \\
 &\times \widetilde\R(\rottheta_{\f})\mathbf{F}(\eta) \widetilde\U_{\text{LMSZ}} %\mathbf{F}(0)^* 
 \widetilde\R(-\rottheta_{\i}) \G \notag \\
 &=
 \G\widetilde\R(\rottheta_{\i}) \widetilde\U_{\text{LMSZ}}^T \mathbf{F}(2\eta) \widetilde\U_{\text{LMSZ}} \widetilde\R(-\rottheta_{\i}) \G,
\end{align}
since $\mathbf{F}(\eta) \mathbf{F}(\eta) = \mathbf{F}(2\eta)$.
The transition probability is $P=|U_{12}(T,0)|^2$; explicitly,
\bse
\begin{align}
\label{P-approx}
P&=\left[\frac{1 + e^{-\pi  \delta ^2}}{2} \sin \left(\eta-2 \chi _1\right) - \frac{1-e^{-\pi  \delta ^2}}{2} \sin \left(\eta -2 \chi _2\right)\right]^2 
 \\
\label{P-approx-2}
&= \left[
 \sin \left(\eta -2 \chi _1\right) 
 - (1-e^{-\pi\delta ^2}) \cos \chi_- \sin \left(\eta -\chi_+\right)
 \right]^2
\\
%or as
\label{P-approx-3}
 &= \left[
\sin \chi_- \cos\left(\eta -\chi _+\right)
-e^{-\pi  \delta ^2} \cos \chi_- \sin \left(\eta -\chi_+\right)
\right] ^2 
\\
\label{P-approx-4}
 &= \left(\sin^2 \chi_- + e^{-2\pi\delta ^2} \cos^2 \chi_-\right) \sin^2 \left(\eta -\chi_+ - \nu \right) ,
\end{align}
\ese
where $\chi_\pm = \chi_1\pm\chi_2$ and
\be
\nu = \arctan(e^{\pi\delta ^2} \tan \chi_-).
\ee
These expressions can be useful in different contexts.
Expression \eqref{P-approx-2} is useful for small $\delta$ when the second term vanishes. 
Expression \eqref{P-approx-3} is useful for large $\delta$ when its second term vanishes.
In particular, on resonance ($\delta=0$) we have $\chi_1 = 0$, $\eta=A/2=\frac12\int_0^T \Omega(t) dt$ and from Eq.~\eqref{P-approx-2} we recover the resonant solution, $P = \sin^2(A/2)$.
The last expression \eqref{P-approx-4} gives explicitly the amplitude (the first factor) and the phase (in the sin$^2$) of the Rabi oscillations.

%%%%%%%%%%%%%%%%%%%%%%%%%%%%%%%%%%%%%%%%%%%%%%%%%%%%%%%%%%%%%%%%%%%%%%%%%%%%%%%%%%%%%%%%%%%%%%%%%%%%%%%%%%%%%%%%%%%%%%%%%%%%%%%%%%%%%%%%%

The expressions for the phases $\chi_1$ and $\chi_2$ can be simplified in the limits of small and large $\delta$.
For small $|\delta|\ll 1$, we have
\bse
\begin{align}
\chi_1 &\sim \frac{\delta ^2}{2} \left[1 -\gamma  -\ln (2 \delta ^2)\right], \\
\chi_2 &\sim \frac{\delta ^2}{2} \left[1 -\gamma  -\ln (\delta ^2/2)\right] -\frac{\pi }{4} , \\
\chi_+ &\sim \delta ^2 \left[1 -\gamma  -\ln (\delta ^2)\right] -\frac{\pi }{4} , \\
\chi_- &\sim \frac{\pi }{4} - \delta^2 \ln 2 , 
\end{align}
\ese
where $\gamma \approx 0.577216\ldots$ is Euler's constant. 

For large $|\delta| \gg 1$, the asymptotic behavior is
\bse\label{chi_large_detun}
\begin{align}
\chi_1 &\sim \frac{1}{12\delta^2}, \\ 
\chi_2 &\sim -\frac{1}{6\delta^2} , \\
\chi_+ &\sim -\frac{1}{12\delta^2} , \\
\chi_- &\sim \frac{1}{4\delta^2} .
\end{align}
\ese
These expressions make it possible to estimate the large-$\delta$ behavior of the transition probability. 
Because the second term in Eq.~\eqref{P-approx-3} vanishes for large $\delta$, the asymptotics is determined by the first term, which behaves as
\be
P\sim \sin^2 \chi_- \cos^2\left(\eta -\chi _+\right)
 \sim \frac{\cos^2\eta}{16\delta^4} .
\ee
Hence the wings of the transition probability vanish as $1/\delta^4$ --- much faster than the wings for the rectangular pulse which vanish as $1/\delta^2$.

Equation \eqref{P-approx-4} allows us to readily estimate the width of the excitation profile, i.e. the value $\Delta_{\frac12}$ at which $P(\Delta)=\frac12$.
Equation \eqref{P-approx-4} describes Rabi oscillations with the first factor being the amplitude and the second factor containing the frequency of the oscillation through the adiabatic phase $\eta$.
The first factor in Eq.~\eqref{P-approx-4} depends on $\delta$ only, whereas the second factor depends also on the adiabatic phase $\eta$. 

We can estimate the linewidth of the excitation profile --- the transition probability versus the detuning $\Delta$ --- by setting the first term in Eq.~\eqref{P-approx-4} to $\frac12$.
The solution is
%\be
$\delta_{\frac12} \approx \pm 3.78$.
%\ee
By recalling the definition of $\delta$ of Eq.~\eqref{delta}, we find
\be
\Delta_{\frac12} T \approx \pm 13.4 \sqrt{\Omega_0 T} .
\label{eq-sin-sqrt}
\ee
We conclude that the line width scales as $\Delta_{\frac12} \propto \sqrt{\Omega_0}$.

Furthermore, the availability of a relatively simple analytic formula allows us to estimate the magnitude of the satellites --- another drawback of the rectangular pulse, which is greatly reduced here.
For a rectangular pulse with a temporal area of $\pi$ and width $T$, the first satellite occurs at detunings $\Delta \approx \pm 8.42/T$ and has a magnitude of 11.64\%.
For comparison, for a sine pulse with a temporal area of $\pi$, the first satellite occurs at detunings $\Delta \approx \pm 11.1/T$ and has a magnitude of just 1.48\%, a reduction by a factor of 8.
This is a consequence of the fact that for large detunings, the transition probability decreases as $1/\Delta^2$ for the rectangular pulse and as $1/\Delta^4$ for the sine pulse~\cite{Boradjiev2013}, as well as for the other pulses of this class considered here, as noted above.
Indeed, this behavior can readily be deduced from Eq.~\eqref{P-approx-4} and the asymptotic behavior of $\chi_-$ in Eq.~\eqref{chi_large_detun}.

It is important to note that
because the oscillation amplitude in Eq.~\eqref{P-approx-4} depends on the parameter $\delta$ only, 
so do the line width and the magnitudes of the sideband peaks.
The parameter $\delta$, as defined in Eq.~\eqref{delta} depends on the detuning $\Delta$ and the maximum Rabi frequency $\Omega_0$ but it does not depend on the particular pulse shape.
The implication is that the conclusions about the line width and the magnitude of the first sideband apply to all pulse shapes, which start and end linearly in time and change smoothly in between.

\section{Demonstration on Superconducting Qubits\label{Sec:demo}}

\subsection{Qubit Specs}

We have conducted series of experiments on the qubit 0 of ibmq\_quito, one of the IBM Quantum Falcon Processors~\cite{ibm}. 
The latter consisted of 5 transmon qubits, and was an open-access quantum processor. 
The exact date of the demonstration was 27 April 2023. 
The parameters of qubit~0 of the ibmq\_quito system at the time of the demonstration were
as follows: the qubit frequency was $5.3007$~GHz, with anharmonicity $-0.3315$~GHz. 
The $T_1$ coherence time was $62.24$~\textit{\textmu}s, the $T_2$ coherence time was $82.71$~\textit{\textmu}s and 
the readout assignment error was $7.14\%$.

The comparison between the excitation landscapes of the sinusoidal and the rectangular pulses was performed on a larger quantum system -- ibm\_kyoto. 
As of taking the measurements on 8 December 2023, the $T_1$ and $T_2$ coherence times were 305.22 and $58.05~\mu$s, while the readout assignment error was $0.29$\%.

The demonstration aims to examine the accuracy of the two derived methods in the context of finite pulses with (nearly) linear side slopes in real context. 
To this end, pulses with multiple envelopes (described in Table~\ref{tab:pulse-shapes}) were applied to the qubit and the post-excitation transition probability was measured and compared to the theoretical predictions.

\subsubsection{Measurement Analytics: Fit Functions and Metrics}

The fits are constructed using the expressions for the transition probability found in Eq.~\eqref{eq-Psplit} by using $P = \left|U_{12}\right|^2$ for the split LMSZ-adiabatic model and Eq.~\eqref{P-approx} for the integrated LMSZ-adiabatic model. 
In addition, a linear correction $P_{\text{final}} = \epsilon_1 + \epsilon_2 (2 P - 1)$ is applied to model the leakage, dephasing, SPAM and readout errors.
The observations are then fitted with the expressions that follow from the two models by  adjusting the free parameters $\ts$, $\epsilon_1$ and $\epsilon_2$, where $0 \le \ts \le T/2$, and $\pm\ts$ are the border points between the rot-LMSZ evolution and the adiabatic evolution. 
The parameter $\ts$ is only relevant for the split LMSZ-adiabatic model. 

The accuracy of the models is analyzed using two metrics: the mean absolute error~(MAE) and the standard deviation of the resonant frequency~(SDRF).
The MAE measures the agreement of the data with the fit, while the SDRF shows the error in determining the resonant frequency.

\begin{table}[]
\begin{tabular}{|c|cr|cc|cc|}
\hline
\multicolumn{1}{|r|}{} & \multicolumn{2}{c|}{Type of Error}                                                     & \multicolumn{2}{c|}{MAE($\text{E}$-3)} & \multicolumn{2}{c|}{SDRF(kHz)}         \\ \hline
Pulse Shape            & \multicolumn{1}{c|}{Rabi Freq.}                  & Model                 & \multicolumn{1}{c|}{Split}   & Int.   & \multicolumn{1}{c|}{Split} & Int. \\ \hline
Sine                   & \multicolumn{2}{c|}{$f(t) = \sin{\left(\frac{\pi t}{T}\right)}$}                       & \multicolumn{1}{c|}{5.5}     & 6.0          & \multicolumn{1}{c|}{33.6}  & 41.7       \\ \hline
Lorentzian             & \multicolumn{2}{c|}{$f_0(t) = \left(1+\left(\frac{t}{\width}\right)^2\right)^{-1}$}   & \multicolumn{1}{c|}{5.5}     & 6.4          & \multicolumn{1}{c|}{37.7}  & 48.1       \\ \hline
Lorentzian$^2$         & \multicolumn{2}{c|}{$f_0(t) = \left(1+\left(\frac{t}{\width}\right)^2\right)^{-2}$} & \multicolumn{1}{c|}{5.9}     & 7.6          & \multicolumn{1}{c|}{34.7}  & 55.1       \\ \hline
Sech                   & \multicolumn{2}{c|}{$f_0(t) = \operatorname{sech}{\left(\frac{t}{\width}\right)}$}     & \multicolumn{1}{c|}{6.2}     & 6.1          & \multicolumn{1}{c|}{40.9}  & 45.1       \\ \hline
Sech$^2$               & \multicolumn{2}{c|}{$f_0(t) = \operatorname{sech}^2{\left(\frac{t}{\width}\right)}$}   & \multicolumn{1}{c|}{5.7}     & 6.0          & \multicolumn{1}{c|}{37.5}  & 42.0       \\ \hline
Gaussian               & \multicolumn{2}{c|}{$f_0(t) = \exp{\left(-\left(\frac{t}{\width}\right)^2\right)}$}    & \multicolumn{1}{c|}{6.2}     & 6.5          & \multicolumn{1}{c|}{34.8}  & 40.4       \\ \hline
\end{tabular}
\caption{This table presents the raw Rabi frequency envelope $f_0$ for each pulse type before the background subtraction and the amplitude rescaling. For the sine pulse the pulse envelope is presented in its final form. The mean absolute errors~(MAE) and standard deviations of the resonant frequency~(SDRF) are also shown for the 6 shapes, obtained with pulse duration equal to the pulse width $\width = 42.67~$ns.}
\label{tab:pulse-shapes}
\end{table}
\subsubsection{Pulses}
\label{sec-pulses}

In order to test the relevance of the two approaches presented in Secs.~\ref{Sec:derivation1} and~\ref{Sec:derivation2}, a set of 6 pulse shapes was selected: sine, Lorentzian, Lorentzian$^2$, sech, sech$^2$ and Gaussian (see Table~\ref{tab:pulse-shapes}). 
The sine function has natural endings and the pulse could be limited to the interval from $0$ to the duration $T$, so the pulse has its endpoints at zero. 
In comparison, the model demands that any infinite pulses are truncated at carefully chosen locations. 
We truncate the shapes at times $0$ and $T$ so that the pulse is centered at $t=\frac{1}{2}T$, and then we subtract the background and rescale the amplitude.

All pulses that were used in this work, both to test the accuracy of the two models and to demonstrate scenarios outside the scope of the models, were of the same pulse width $\width$. 
They included the pulse shapes in Table \ref{tab:pulse-shapes}, as well as the exponential shape $f_0(t) = \exp{\left(-\left|t/\width\right|\right)}$.
Because of the equal pulse widths, their transition line profiles owed the differences in their characteristics mainly to the shape of the pulses. 
The two approximations assume linearity in the initial and final segments of the pulse, as stated in Sec.~\ref{Sec:derivation2}.
This condition is met closely for the sine function, since the first-order term in its Taylor series at $t=0$ and $t=T$ is linear in $t$.
The other 5 pulses were also truncated at the same duration $T=\width$ to analyze the applicability of the models.

The tests used a pulse width of $\tau=42.67$~ns for all continuous pulses in the study.
For the sine pulse this value corresponds to half a period, where the pulse is halted. 
The other models' duration values were equal to their pulse width $\width$, so as to truncate close to the points of steepest gradient.
Note that the points of steepest gradient stand at the zeroes of the second derivative of the pulse shape; hence at this point of time the pulse shape is as linear as it gets.
The area of the pulse envelopes is calibrated prior to the tests to be equal to $\pi$. 
The Qiskit framework only reports the value of the Rabi frequency in arbitrary units. 
Thus, the value in SI units was calculated using the calibrated pulse area and the envelope. 

Furthermore, a study of the limitations of the two models was made using two pulse shapes: Lorentzian$^2$, to demonstrate what happens when slopes are non-linear and the linearity assumption is not fulfilled, and exponential, to analyze what happens when the adiabatic assumption is broken in the central part of the pulse.
A duration value of $1.67\width$ was used for the Lorentzian$^2$ pulse such that the effects from increasing non-linearity could be compared with the Lorentzian$^2$ pulse with duration $\width$. 
The exponential pulse profile was demonstrated with a single duration $T=\width$.

The amplitude parameter in Qiskit is used to control the Rabi frequency of the driving field and allows the application of pulsed excitation on the qubit.
However, its relation to the Rabi frequency is not linear, and lags behind with greater amplitude values.
Therefore, a mapping is used to match the amplitude values to the actual Rabi frequency of the field, using a fit with a custom function.

\subsection{Results and Discussion}

\begin{figure}[tb]
    \centering 
    \includegraphics[width=\columnwidth]{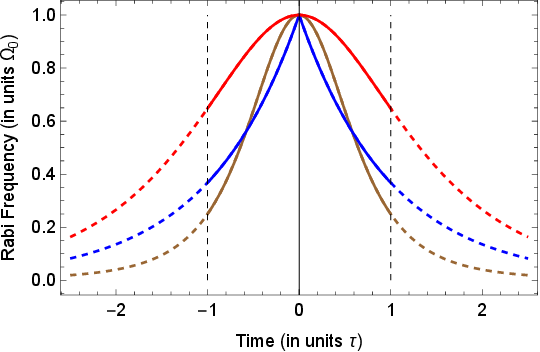}
    \caption{(Color online) Sech~(red), exponential~(blue) and Lorentzian$^2$~(brown) pulse envelopes with the same pulse width $\width$. The dotted vertical lines show the truncation line scaled for the pulses with equal pulse width and duration, which is the case for most pulses used in this demonstration.}
    \label{fig:truncation}
\end{figure}

\subsubsection{Sine pulse}

\begin{figure}[tb]
    \centering
    \includegraphics[width=\columnwidth]{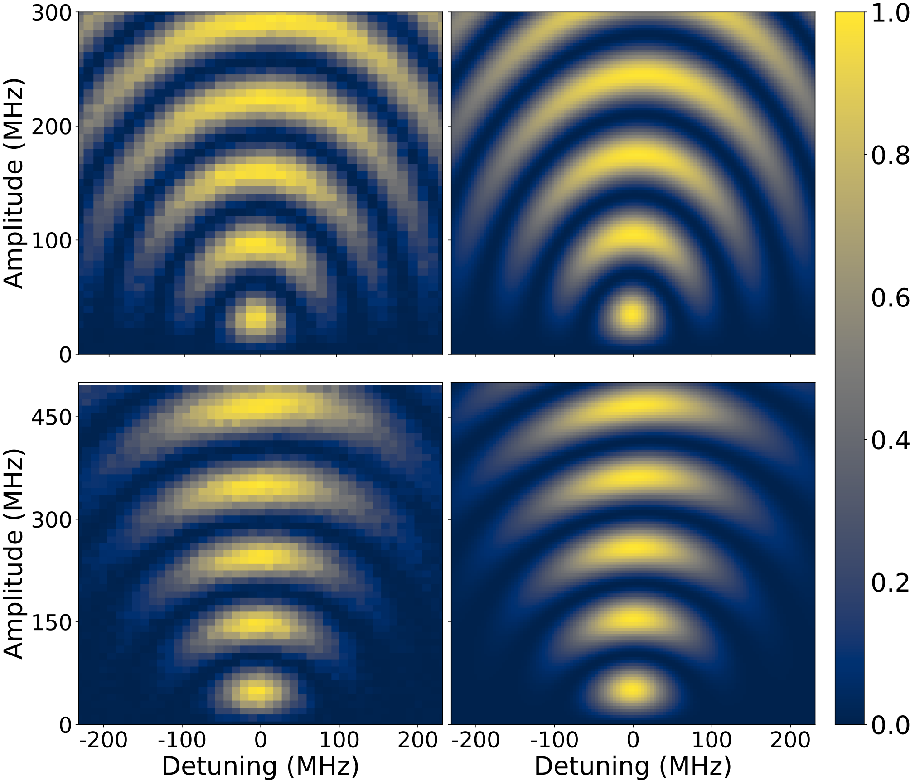}
    \caption{(Color online) The observed (left) and numerical (right) excitation landscapes (transition probability against detuning and peak Rabi frequency) of rectangular (top) and sinusoidal (bottom) pulsed excitation enable us to compare the transition linewidth power expansion of the two models.}
    \label{fig:broadening}
\end{figure}

The tests of the model applicability were preceded by a detailed demonstration of its relevance.
We begin with the sine pulse  in its role of a prime example of the type of pulses studied here.
The Taylor decomposition of the sine function around the $t=0$ point shows linear evolution in the first order, with zero second order contribution, making it the perfect case studies.
Furthermore, the natural ending points at sine arguments $0$ and $\pi$ suggest no need for post-processing (truncation, background subtraction and amplitude rescaling), which is required for the other pulse shapes.

\begin{figure}[tb]
    \centering
    \includegraphics[width=\columnwidth]{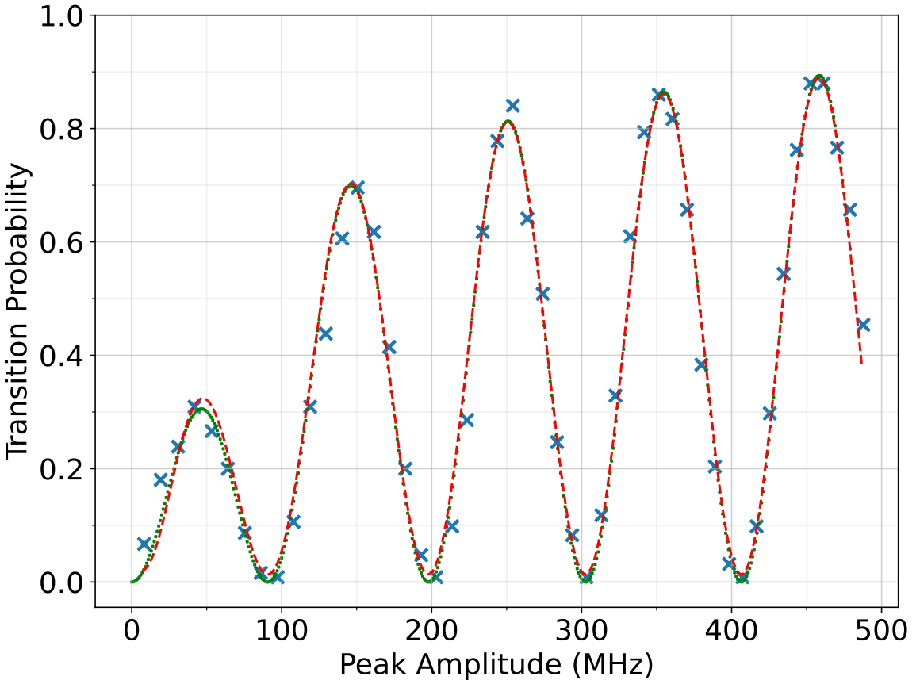}
    \caption{(Color online) Measurements of off-resonant Rabi oscillations of a sine pulse, with the same width $\width = 96.0$~ns as in Fig.~\ref{fig:broadening}, at detuning $\Delta=-46.2$~MHz~(blue crosses) fitted with the analytical expression for the sine model found using the integrated model~(red dashed line), and with the numerical solution~(green dotted line). }
    \label{fig:rabiosc}
\end{figure}

The excitation generated by a sine pulse was measured and compared against that of a rectangular pulse.
Demonstrations of the excitation landscapes of rectangular and sinusoidal pulses with widths of $96.0$~ns were performed on IBM Quantum's ibm\_kyoto processor. 
The results are displayed in Fig.~\ref{fig:broadening}.
We identify the first 5 maxima of the transition probability (corresponding to temporal pulse areas of $\pi$, $3\pi$, $\hdots,$ $9\pi$) in the excitation landscapes.
The figure shows that sinuisoidal pulsed excitation delivers a more compact, less power-broadened transition than the rectangular pulse while sustaining its finite span. 
The transition line width is proportional to the square root of the driving amplitude for the sine pulse [see Eq.~\eqref{eq-sin-sqrt}], compared to the linear dependence in the rectangular pulse. 
This signals a quadratic decrease in power broadening effects.

The off-resonant Rabi oscillations of a sine pulse are shown in Fig.~\ref{fig:rabiosc}. 
Generally, the plot indicates good agreement between the measurements on the quantum hardware and the integrated model predictions. 
Nevertheless, minor inconsistencies between them are present, especially around the first maximum.
It is evident that the experimental points consistently are delayed on the left-hand side of this figure and then begin to overtake the theoretical prediction toward the right-hand side.
We attribute this minor discrepancy to the non-linearity of the Qiskit amplitude parameter with respect to the Rabi frequency (mentioned at the end of Sec.~\ref{sec-pulses}), which introduces a non-negligible error on the Rabi oscillations.

It is evident in Fig.~\ref{fig:broadening} that the measured excitation landscapes~(left column) are slightly skewed to the right -- in the responses to both the rectangular and the sine pulse.
We note that such asymmetry might be an indication of leakage.
Indeed, while the closed qubit dynamics is symmetric to the detuning, adding more states to the qubit makes excitation asymmetric versus the detuning, which can also be explained in terms of an induced ac Stark shift due to the presence of higher levels.

\begin{figure}[tb]
    \centering
    \includegraphics[width=\linewidth]{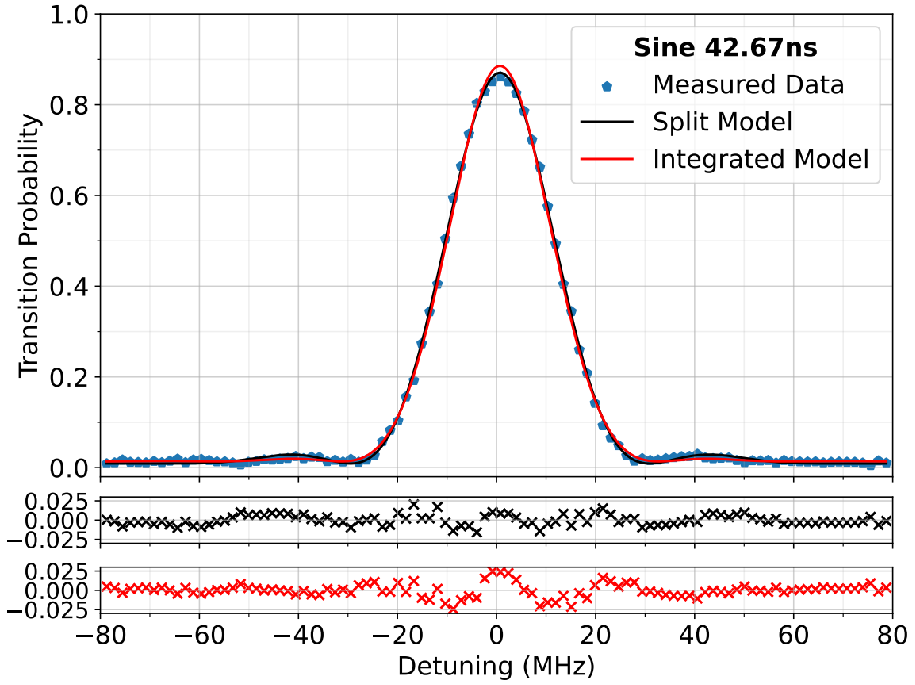}
    \caption{(Color online) The measured transition line profile of the sinusoidal pulse envelope~($T=42.67$~ns) is plotted~(blue hexagons) along with fits based on the split approximation~(in red), and the integrated approximation~(in green). The residuals of the two models are shown in the thin axes below the plot.
%    \textcolor{red}{\sc Kolko e $\ts$?}
}
    \label{fig:sine}
\end{figure}

\begin{figure*}[tbh]
    \centering
    \includegraphics[width=\linewidth]{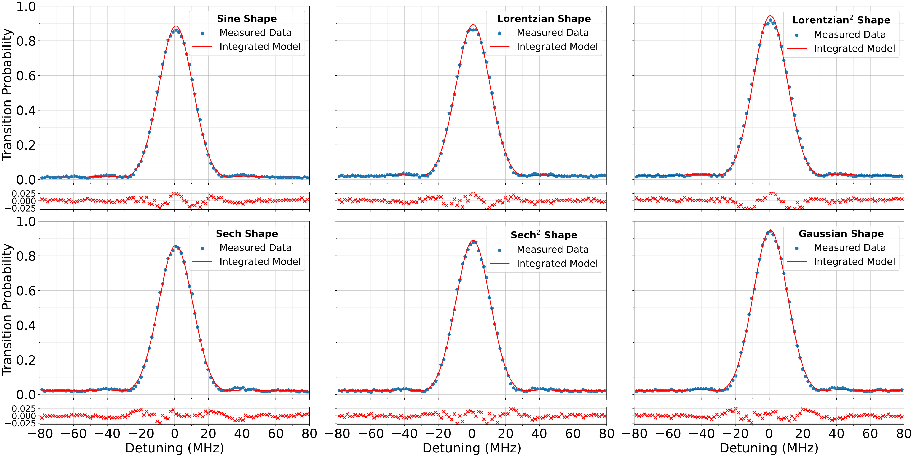}
    \caption{(Color online) The measured transition line profiles~(blue hexagons) is plotted along with fits based on the integrated approximation~(in green) for 6 pulse shapes found in Table~\ref{tab:pulse-shapes} with $\width=T=42.67$~ns. Top row: (left to right) Sine, Lorentzian, Lorentzian$^2$; Bottom row: Sech, Sech$^2$ and Gaussian. The residuals of the two models are shown in the thin axes below each plot.}
    \label{fig:dur196}
\end{figure*}

The two approximations derived in this work were tested against the transition line profile of a sinusoidal pulse with Rabi frequency $\Omega_0 = 115.7$~MHz and pulse width $42.67$~ns.
The transition line profile acquired on the ibmq\_quito system is presented in Fig.~\ref{fig:sine}, along with fits based on the two models.
A brief analysis of the residuals at the bottom of Fig.~\ref{fig:sine} reveals general agreement between the two approximate models. 
Furthermore, they reflect the behavior of the data, following its pattern closely even in the neighborhood of the sensitive satellite peaks at $\Delta \approx \pm 40$~MHz.
The MAE values of the two fits set a standard against which the compatibility of the models with other candidate pulses can be evaluated. 
The two models deliver similar accuracy when applied to the sine model.
The integrated model results in a 9\% increase of the MAE and 24\% increase of the SDRF when compared to the split method. 
Because the two approximations deliver comparable results, despite their very different appearances, we shall use hereafter the simpler, integrated approximation only.

\subsubsection{Other pulse shapes of finite duration}

Similar demonstrations on a total of six pulse shapes (incl. the sine pulse) are presented in Fig.~\ref{fig:dur196}, and their performance indicators are listed in Table~\ref{tab:pulse-shapes}.
In order to satisfy the assumptions of the analytic models, the continuous waveform shapes were symmetrically truncated on the steep segments of their slopes, as shown in Fig.~\ref{fig:truncation}. 
The resulting additive background was subtracted, so that the pulses begin and end at zero Rabi frequency.
The measured transition probability for the six pulses was then fit using the integrated model.
The data for each of the pulses can be examined in Fig.~\ref{fig:dur196}, where the corresponding residuals of the fits are shown underneath each of the plots.
The examination of the transition line profiles of the 6 pulses reveals that the accuracy of the approximation is sufficient when the two principal assumptions of the models are satisfied by the pulse shape. 
The central assumptions of the models are outlined in Secs.~\ref{Sec:general}, \ref{Sec:derivation1} and \ref{Sec:derivation2}: linearity of the slopes at the beginning and end of the pulse, and adiabaticity of the pulse around its peak.
It is quite remarkable that the plots look very similar, including the line widths and the small sidebands, which confirms the assertion that these are governed by the parameter $\delta$ of Eq.~\eqref{delta}.

\section{Validity of the model\label{Sec:applicability}}

The range of pulse shapes where the two models are applicable was analyzed with the help of the Lorentzian$^2$ and the exponential shapes. 
They both violate the initial assumptions of the models, albeit in a different manner.
The Lorentzian$^2$ pulse's initial and final slopes can be made less linear by moving the truncation point away from the steepest decent of the shape, thereby introducing nonlinearities to it. 
In so doing both $\dot\rottheta(t)$ decreases and $\ddot\rottheta(t)$ increases, thereby rendering the linearity assumption break down.
The exponential shape violates the other condition: that the pulse is nearly adiabatic in the vicinity of the central pulse summit.

\begin{figure}[tbh]
  \centering
\includegraphics[width=\linewidth]{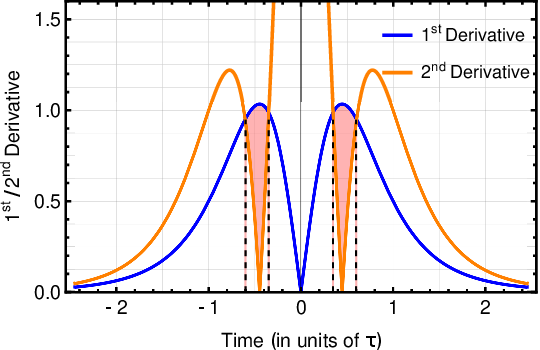}
\caption{(Color online) The magnitudes of the first and second derivatives of a Lorentzian$^2$ pulse are shown and the interval where the pulse is approx. linear is highlighted in red.}
\label{fig:lor2-der}
\end{figure}

\subsection{Breaking the linearity condition}

The suitability of the models for a given pulse shape can be estimated before experiment, see Fig.~\ref{fig:lor2-der}. 
The linearity requirement can be estimated using different metrics and evaluated against a threshold. 
Likewise, the second derivative of its Rabi frequency $\Omega(t)$ in the vicinity of the truncation point can be compared against an upper limit on its magnitude to check the appropriateness of the model beforehand.
In general, the most optimal truncation point is in the vicinity of the inflection point where both the second derivative is zero and the first derivative is maximal in magnitude.

\begin{figure}[tbh]
\centering
  \includegraphics[width=\linewidth]{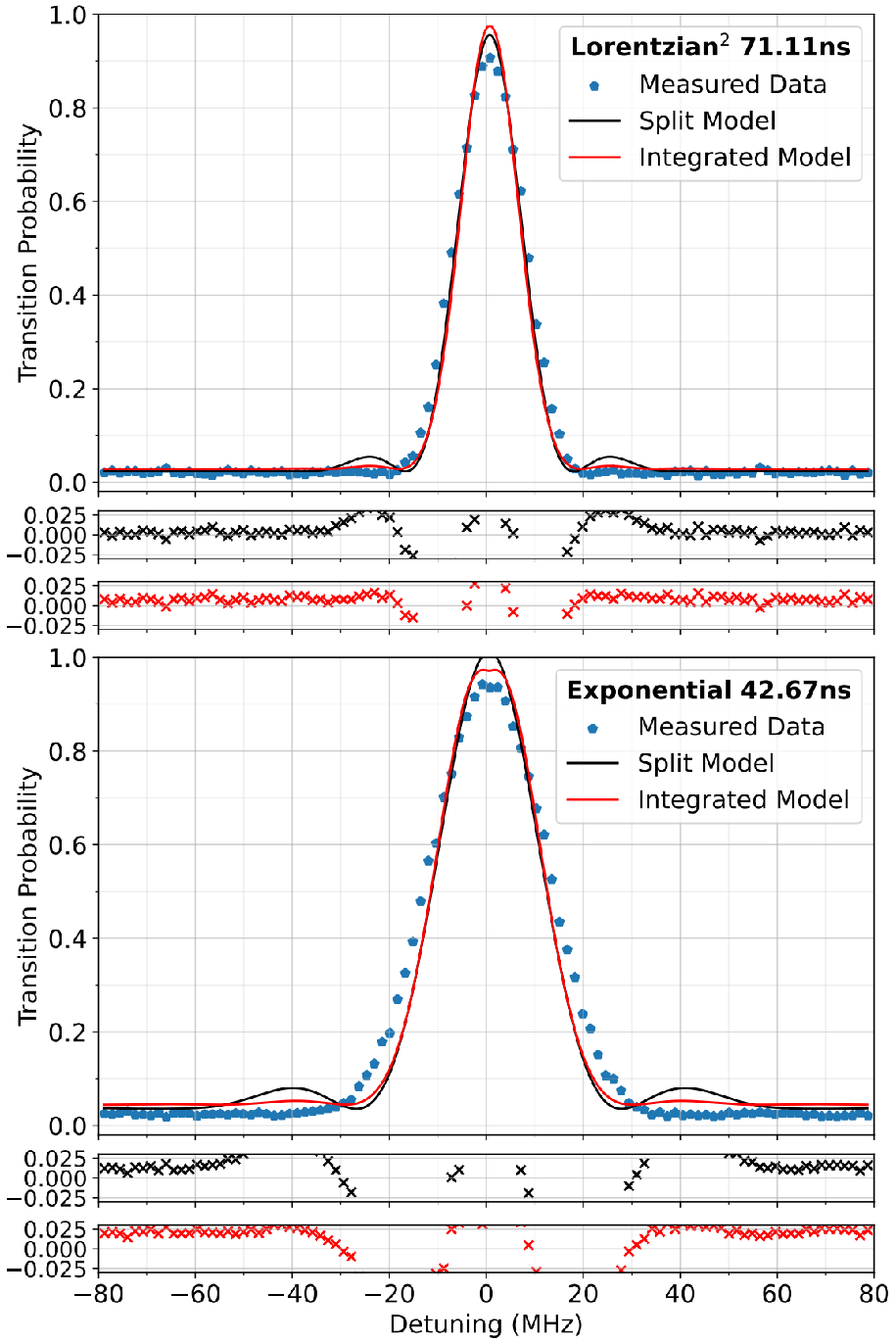}
  \caption{(Color online) The measured transition line profiles of the Lorentzian$^2$ (top) and exponential (bottom) pulse envelopes with durations -- $71.11$~ns and $42.67$~ns respectively -- are plotted~(blue hexagons) along with fits based on the split~(in black), and the integrated method~(in red). }
  \label{fig:failures}
\end{figure}

However, displacing the truncation points would immediately be reflected in the error of the fit, because of the non-linearity that would be present close to the ends of the pulse.
The Lorentzian$^2$ pulse has a steep slope, which makes it highly susceptible to movement of the truncation points.
This is evident from Fig.~\ref{fig:failures} where the Lorentzian$^2$ pulse from Fig.~\ref{fig:dur196} was truncated further from its inflection point, and its duration is $71.11$~ns, or 1.67$\width$, compared to $42.67$~ns at the inflection point (see Fig.~\ref{fig:dur196}).
Figure~\ref{fig:failures} exhibits some discrepancies between theory and experiment both in the central part of the profile and in the predicted but unobserved sidebands.
These issues are reflected in the MAE and SDRF of the fit, which increase in comparison with the shorter-duration Lorentzian$^2$ pulse on Fig.~\ref{fig:dur196}.
For the longer Lorentzian$^2$ pulse (Fig.~\ref{fig:failures}), the split and integrated models have MAE's of $15.7 \times 10^{-3}$ and $17.8 \times 10^{-3}$ and SDRF's of $87.3$ and $99.4$~kHz respectively, which signal a three-fold increase of the error, compared to the $T=42.67$~ns Lorentzian$^2$ pulse in Fig.~\ref{fig:dur196}.

These values are consistent with the truncation location analysis, as shown in Fig.~\ref{fig:lor2-der}.
It is evident that the second derivative of the Lorentzian$^2$ pulse is smaller than its first derivative for durations $T\in(0.7\width, \width)$, where the ends lie in the vicinity of the inflection points.
Notably, the integrated model is comparable in predicting the behaviour of the Lorentzian$^2$ pulse to the split model.
It is likely that this is because the adiabatic phase (see Eq.~\eqref{eq-IntAdPhase}) is taken exactly, where the split model approximates it as that of a linear coupling in the time interval $[0,\ts]$.
Nevertheless, the assumption of slope linearity is still applied in the integrated model to find the propagator in Eq.~\eqref{ULMSZ}.

\subsection{Breaking the adiabatic condition}

The adiabatic condition can similarly be calculated for the corresponding pulse shape.
However, one should take into account that the location where the adiabatic condition is most unstable depends on the exact shape of the pulse.
For example, the exponential pulse, shown in Fig.~\ref{fig:truncation}, has a discontinuity in the derivative exactly at its tip. 
This suggest that the point of highest non-adiabaticity is its central point. 
In contrast, pulses with smooth peaks are usually least adiabatic on their slopes, close to the inflection points, where their first derivatives have high magnitude values. 
In the limiting case of the exponential model, presented in the bottom plot of Fig.~\ref{fig:failures}, the apparent non-adiabatic behaviour at the central point significantly disrupts the applicability of the models.
It is the outlier in this demonstration. 
Regardless of the highly linear slopes at the ends of the exponential pulse, the non-adiabatic behavior at its summit ruins the fits.
The MAE and SDRF errors exceed these of the other pulses by factors between 5 and 8. 
To this end, we note that as far as the exponential model is concerned, one can derive an analytic approximation by adapting the derivation of the Demkov model \cite{Demkov1963} for a finite-duration pulse.

\section{Conclusions\label{Sec:conclusion}}

In this paper we have presented comprehensive theoretical description and experimental demonstration of the dynamics of a qubit driven by a pulse of finite duration with linearly ending wings.
Such pulse shapes, the most prominent member of which is the sine pulse between two of its nodes, hold some advantages over both the rectangular pulse and the truncated pulses of otherwise infinite duration, such as Gaussian. 
Compared to the former, the present pulse shape exhibit much less power broadening and an order of magnitude smaller sidebands. 
Compared to the latter, the present pulse shape has a well defined finite duration and avoids truncation artefacts, e.g. truncation-induced power broadening.

We have derived two analytic approximations in order to describe the excitation dynamics.
Both of them make use of a rotated picture wherein the problem resembles two LMSZ half-crossings separated by smooth adiabatic evolution. 
One of these, the split approximation, splits the evolution into a half crossing, adiabatic evolution and another half crossing and gives the solution in terms of Weber's parabolic cylinder function.
The other, integrated approximation, uses the asymptotics of the Weber's function and integrates the LMSZ and adiabatic evolution within the adiabatic interaction representation, where the LMSZ dynamics is localized in the vicinity of the half crossings. 
The integrated approximation delivers a much simple analytic formula which allows one to draw important conclusions about the excitation profile features.
Among these, we explicitly find square-root power broadening, $\Delta_{\frac12} \propto \Omega_0^{\frac12}$, the $\Delta^{-4}$ law for the amplitude of the excitation profile, and an order of magnitude smaller sidebands compared to the rectangular pulse.
Moreover, the integrated approximation shows that the transition probability depends primarily on the slope of the pulse shape in the turn-on and turn-off times and the adiabatic phase over its course, but not so much on the specific pulse shape.

By comparing the two approximations to the experimental data acquired on two of IBM Quantum processors we have found an excellent agreement between theory and experiment in all cases when the linear and adiabatic conditions have been fulfilled.
We have exploited 6 different pulse shapes which show very similar excitation profiles, thereby verifying the analytic findings that, as noted above, the specific pulse shape is of no great significance as long as the pulse shape turns on and off linearly and lacks sharp features in between. 
We have also studied pulse shapes which violate either the linearity or the adiabaticity conditions in order to find out the limitations of the analytic approximations.

%%%%%%%%%%%%%%%%%%%%%%%%%%%%%%%%%%%%%%%%%%%%%%%%%%%%%%%%%%%%%%%%%%%%%%%%%%%%%%%%%%%%%%%%%%%%%%%%%%%%%%%%%%%%%%%%%%%%%%%%%%%%%%%%%%%%%%%%%%%%%%%%%%%%%%%%%%%%%%%%
\acknowledgments

We gratefully acknowledge the Karoll Knowledge Foundation for providing financial support to Ivo S. Mihov during the preparation of this manuscript. Their contribution was invaluable in encouraging the author's ongoing scientific endeavours. This research is supported by the Bulgarian national plan for recovery and resilience, Contract No. BG-RRP-2.004-0008-C01 (SUMMIT), Project No. 3.1.4 and by the European Union’s Horizon Europe research and innovation program under Grant Agreement No. 101046968 (BRISQ).
We acknowledge the use of IBM Quantum services for this work. 
The views expressed are those of the authors, and do not reflect the official policy or position of IBM or the IBM Quantum team.

\appendix

%%%%%%%%%%%%%%%%%%%%%%%%%%%%%%%%%%%%%%%%%%%%%%%%%%%%%%%%%%%%%%%%%%%%%%%%%%%%%%%%%%%%%%%%%%%%%%%%%%%%%%%%%%%%%%%%%%%%%%%%%%%%%%%%%%%%%%%%%

%%%%%%%%%%%%%%%%%%%%%%%%%%%%%%%%%%%%%%%%%%%%%%%%%%%%%%%%%%%%%%%%%%%%%%%%%%%%%%%%%%%%%%%%%%%%%%%%%%%%%%%%%%%%%%%%%%%%%%%%%%%%%%%%%%%%%%%%%
\section{Half LMSZ propagator in the adiabatic interaction representation\label{Sec:AdbInt-LMSZ}}

For the LMSZ model, the propagator reads
\be\label{Cayley-Klein}
\left[ \begin{array}{cc} a & b \\ -b^* & a^* \end{array} \right]
\ee
where the Cayley-Klein parameters are given by Eq.~\eqref{ab-LMSZ}.
We will use the so-called strong-coupling asymptotics of Weber's parabolic cylinder functions ($\alpha^2+\delta^2 \gg 1$),
\bse
\begin{align}
D_{-i \delta ^2}(\alpha  e^{\frac{i \pi }{4}}) &\sim \cos \zeta  \exp \left(\frac{\pi \delta^2}{4} - i\etaLMSZ \right),\\
D_{i \delta ^2-1}(\alpha  e^{-\frac{i \pi }{4}}) &\sim \frac{\sin \zeta }{\delta }  \exp \left(\frac{\pi \delta^2}{4}+\frac{i\pi}{4} + i\etaLMSZ \right),
\end{align}
\ese
where
\be
\zeta = \frac12 \arctan \frac {2\delta}{\alpha} .
\ee
and
\be\label{etaLMSZ}
\etaLMSZ = \frac{\alpha}{2}  \sqrt{\alpha ^2 + 4\delta ^2} + 2\delta ^2 \ln \left(\frac{\sqrt{\alpha ^2 + 4\delta ^2}+\alpha }{2}\right) -\delta ^2 \ln\delta^2 .
\ee
We note that $\etaLMSZ$ is exactly the adiabatic phase for this model,
\be
\etaLMSZ = \int_0^{\frac12 T} \sqrt{\alpha^2 t^2 + 4\delta^2}\, dt.
\ee

%%%%%%%%%%%%%%%%%%%%%%%%%%%%%%%%%%%%%%%%%%%%%%%%%%%%%%%%%%%%%%%%%%%%%%%%%%%%%%%%%%%%%%%%%%%%%%%%%%%%%%%%%%%%%%%%%%%%%%%%%%%%%%%%%%%%%%%%%

By using these asymptotics we find
\bse\label{ab}
\begin{align}
a &= e^{i \xi _1} \sqrt{\frac{1+e^{-\pi\delta^2}}{2}} \cos \zeta
+ e^{-i \xi _2} \sqrt{\frac{1-e^{-\pi\delta^2}}{2}} \sin \zeta\, 
b &= e^{i \xi _2} \sqrt{\frac{1-e^{-\pi\delta^2}}{2}} \cos \zeta\, 
- e^{-i \xi_1} \sqrt{\frac{1+e^{-\pi\delta^2}}{2}} \sin \zeta,
\notag\\
\end{align}
\ese
where
\be
\xi _1 = \chi_1 -\frac{\etaLMSZ}{2} ,\quad
\xi _2 = \chi_2 - \frac{\etaLMSZ}{2} ,
\ee
where $\chi_1$ and $\chi_2$ are defined by Eqs.~\eqref{chi1-chi2}.
We can write the propagator of Eq.~\eqref{Cayley-Klein} with $a$ and $b$ from Eqs.~\eqref{ab} as (accounting for $t_{\i}=0$ and $t_{\f} = T/2$)
\be
\U = \R(-\zeta) \U_{\text{adb}} \U_{\text{LMSZ}},
\ee
where
\be
\R(-\zeta) = 
\left[
\begin{array}{cc}
 \cos \zeta & -\sin \zeta \\
 \sin \zeta & \cos \zeta
\end{array}
\right],
\ee
\be
\U_{\text{adb}} = 
\left[
\begin{array}{cc}
 e^{-\frac{1}{2} i \etaLMSZ} & 0 \\
 0 & e^{\frac12 i \etaLMSZ}
\end{array}
\right],
\ee
\be\label{U_LMSZ_ai}
\U_{\text{LMSZ}}=
\left[
\begin{array}{cc}
 q & s \\
 -s^* & q^*
\end{array}
\right],
\ee
with
\be
q = \sqrt{\frac{1+e^{-\pi\delta^2}}{2}} e^{i \chi _1},\quad
s = \sqrt{\frac{1-e^{-\pi\delta^2}}{2}} e^{i \chi _2}. %\text{sgn}(\delta ).
\ee
Equation \eqref{U_LMSZ_ai} represents the half-LMSZ propagator in the adiabatic interaction representation.

%%%%%%%%%%%%%%%%%%%%%%%%%%%%%%%%%%%%%%%%%%%%%%%%%%%%%%%%%%%%%%%%%%%%%%%%%%%%%%%%%%%%%%%%%%%%%%%%%%%%%%%%%%%%%%%%%%%%%%%%%%%%%%%%%%%%%%%%%%%%%%%%%%%%%%%%%%%%%%%%
%%%%%%%%%%%%%%%%%%%%%%%%%%%%%%%%%%%%%%%%%%%%%%%%%%%%%%%%%%%%%%%%%%%%%%%%%%%%%%%%%%%%%%%%%%%%%%%%%%%%%%%%%%%%%%%%%%%%%%%%%%%%%%%%%%%%%%%%%%%%%%%%%%%%%%%%%%%%%%%%
%%%%%%%%%%%%%%%%%%%%%%%%%%%%%%%%%%%%%%%%%%%%%%%%%%%%%%%%%%%%%%%%%%%%%%%%%%%%%%%%%%%%%%%%%%%%%%%%%%%%%%%%%%%%%%%%%%%%%%%%%%%%%%%%%%%%%%%%%%%%%%%%%%%%%%%%%%%%%%%%

\end{document}